\documentclass{ws-ijmpa}
\newcommand{\be}{\begin{equation}}
\newcommand{\ee}{\end{equation}}
\newcommand{\bear}{\begin{eqnarray}}
\newcommand{\eear}{\end{eqnarray}}
\begin{document}

%\markboth{Authors' Names}
%{}

\catchline{}{}{}{}{}
%
%%%%%%%%%%%%%%%%%%%%%%%%%%%%%%%%%%%%%%%%%%%%%%%%%%%%%%%%%%%%%%%%%%%%

\title{CHIRAL RADIATIVE CORRECTIONS IN POTENTIAL MODEL AND THE
P-WAVE STATES OF HEAVY-LIGHT MESONS}

\author{TAEKOON LEE}

\address{\it Physics Department, Kunsan National University, 
Kunsan 573-701,\\ Korea\\
tlee@kunsan.ac.kr}

\maketitle

%\begin{history}
%\received{Day Month Year} 
%\revised{Day Month Year}
%\end{history}

\begin{abstract}
In relativistic potential model of 
heavy-light mesons  it is essential to incorporate 
the chiral radiative corrections before comparing the
model predictions with  data. Once the radiative corrections
are taken into account the potential model can explain 
the unusual mass spectra and absence of the spin orbit-inversion
in P-wave states.

\end{abstract}

\ccode{PACS numbers: 12.39.Pn, 14.40.Lb, 14.40.Nd}

\section{Introduction}	
It was noted from the very moment of the first observation of a
P-wave state of heavy-light mesons that the relativistic potential
model was in trouble in explaining the data ~\cite{barbar,belle}.
For instance, the
mass of $D_s(2317)$ was too low compared to its nonstrange counterpart
$D(2308)$ and because of this the potential model could not
explain the observed mass gap
\bear
{\rm gap}\equiv 
[m(D(0^+))-m(D(0^-))]-[m(D_s(0^+))-m(D_s(0^-))] \approx 95\,\, {\rm MeV}
\eear
which in the potential model is virtually vanishing.

Another puzzle is the absence of 
the spin-orbit inversion in P-wave states that has been a generic
prediction of potential models.
Schnitzer suggested long ago that the strong spin-orbit
interaction of the scalar confining potential would lead to spin-orbit
inversion in P-wave heavy-light mesons, with the claim
that its observation would confirm the scalar nature of the 
confining potential ~\cite{schnitzer}. This spin-orbit inversion
was later reaffirmed in studies with  more sophisticated  potential
models ~\cite{inversion1,inversion2}.
However, contrary to these studies,  the observed
masses of the P-wave charmed mesons
do not exhibit  spin-orbit inversion.

Because of the apparent failure of the potential model
various alternatives were suggested ~\cite{alternatives}.
It is worth noting, however, that the conventional potential model
had not treated  the
chiral loop corrections consistently.
The radiative correction to the mass of a resonance is given by
the real part of the self energy, and this was not taken into
account in the calculation of the mass spectra, while the
imaginary part was used in computing the decay width. Since the
widths are generally a few hundred MeVs it can be seen immediately
that the loop corrections to the mass spectra cannot be ignored.

In this note we review the chiral loop corrections in the
potential model and show that the puzzles of the P-wave states
can be answered within the potential model once the
loop corrections are taken into account.

\section{Chiral Radiative Corrections}

The one loop 
 corrections for S and P-wave states
 with $j=\frac{1}{2}$ to
 estimate the mass gap was computed in Ref.~\cite{lee1}
 and the calculation was extended to P-wave 
 states with $j=3/2$ as well as D-wave states
 with $j=3/2$ and $ 5/2$ in Ref. ~\cite{lee2} to solve 
 the spin-orbit inversion problem.
 
 The loop corrections in the potential model which is based on the
 chiral quark model are UV divergent and in Refs. ~\cite{lee1,lee2}
 a three-momentum
 cutoff regularization was used. While the loop corrections to the
 energy levels are cutoff dependent the mass gap, which depends on the
 mass differences of the resonances, is virtually independent on the
 cutoff since only the low energy fluctuations of $k\approx 250$ MeV 
 contribute to it. 

 With the UV cutoff at 700 MeV the loop corrections for the energy levels for
 S and P-wave
 states are given as in Table ~\ref{tab.loop}.
\begin{table}[h]
\tbl{Loop corrections $\delta E^{\rm loop}_{l,j,q}$ for S
and P-wave states
in the lowest radial excitations. (Units are in MeV.)}
{\begin{tabular}{@{}cccccccc@{}}\toprule
$\delta E^{\rm loop}_{0,\frac{1}{2},d}$&
$\delta E^{\rm loop}_{0,\frac{1}{2},s}$&
$\delta E^{\rm loop}_{1,\frac{1}{2},d}$&
$\delta E^{\rm loop}_{1,\frac{1}{2},s}$&
$\delta E^{\rm loop}_{1,\frac{3}{2},d}$&
$\delta E^{\rm loop}_{1,\frac{3}{2},s}$\\ \hline
-161&-160&-260&-343&-183&-181 \\ \botrule
\end{tabular}
\label{tab.loop}}
\end{table}

\section{Mass Gap}

We can now see how the mass gap  is affected
by the radiative corrections. For this we focus on the 
effects of the loop
corrections on the existing potential model
predictions in Ref. ~\cite{pe}.

The new energy levels, denoted by $\bar {E}_{\bf m}$, that include 
the radiative
corrections can be related to the energy levels of the conventional
potential model by
\bear
\bar{E}_{\bf m}=E_{\bf m} +\delta E^0_{\bf m} +
\delta E^{\rm loop}_{\bf m}
\label{relation}
\eear
where $E_{\bf m}$ denotes the conventional energy levels which contain
the leading order level $E^0_{\bf m}$ as well 
as the $1/{\rm M}$ corrections, and
$\delta E^0_{\bf m}$ denotes the shift in the
leading order level  $E^0_{\bf m}$
caused by the shift in the fitted values of the parameters of 
the model, which was induced by the introduction of
radiative corrections $\delta E^{\rm loop}_{\bf m}$ in 
the fitting of the
parameters.
The relation  (\ref{relation})
is valid to the
leading order of the loop corrections.

Now noting that the gap for the leading order levels is vanishing we expect
the correction to it to be small and so obtain approximately
\bear
{\rm gap}^{\rm new}\approx{\rm gap}^{\rm old}+ 
{\rm gap}^{\rm loop}=72 \,\,{\rm MeV}
\eear
which is consistent with the experimental value $95$ MeV.

\section{Spin-Orbit Inversion}

Now applying the relation (\ref{relation}) to states with differing  $j$ but
otherwise same quantum numbers in the heavy-quark limit we obtain
approximately 
\bear
\bar{E}_{j}-\bar{E}_{j'}=E_j-E_{j'}
+\delta E^{\rm loop}_{j}-\delta E^{\rm loop}_{j'}\,.
\label{diffrel2}
\eear

Let us  focus on the
spin-orbit inversions in P-wave 
states. 
Looking on Table ~\ref{tab.loop}
we notice that the magnitudes of the loop corrections are 
larger for states with smaller $j$, and this feature
will be crucial for understanding 
the absence of spin-orbit inversions.

We shall first consider the effects of the loop corrections
on the P-wave   $D_s$ mesons. In the following all the states,
labeled  $H(l,j,J)$, are in
their lowest radial excitations. We
shall  assume that the modified energy level for the state
$D_s(1,\frac{1}{2},0)$ coincides with the experimental mass
of $D_s(2317)$, and then estimate
the masses of $j=1/2$ and $3/2$ states. Reading the values
of the conventional energy levels from Ref. ~\cite{pe} and loop
corrections from Table ~\ref{tab.loop} we find 
the following new energy levels of the states 
related by Eq. ~(\ref{diffrel2}):
\bear
\bar E_{D_s(1,\frac{1}{2},1)}&=&2435 \,\,\,{\rm MeV} \,,\nonumber \\
\bar E_{D_s(1,\frac{3}{2},1)}&=& 2527 \,\,\,{\rm MeV} \,,\nonumber \\
\bar E_{D_s(1,\frac{3}{2},2)}&=& 2573 \,\,\,{\rm MeV}\,.
\eear
Comparing this result with the experimental values 2460 MeV,
2535 MeV, and 2573 MeV, respectively, we see there is good 
agreement between the
new levels and data, and there are no longer  spin-orbit inversions. 

Using the same procedure  we can obtain the modified energy
levels for P-wave  $D$ mesons as well, and the result is summarized in
Table ~\ref{c-mesons}.

We can  now use Eq. ~(\ref{relation})
to estimate the masses of the P-wave bottom mesons.
We shall first compute the mass for $B_s(1,\frac{1}{2},0)$, which is
the  counterpart of $D_s(2317)$. 
Since the loop corrections are independent of the heavy quark mass we see that
the last two terms in Eq. ~(\ref{relation}) to be heavy-quark mass
independent, so
we get
\be
\bar{E}_{B_s(l,j,J)}-E_{B_s(l,j,J)}=\bar{E}_{D_s(l,j,J)}-E_{D_s(l,j,J)}\,.
\label{charmbottomrel}
\ee
Identifying again $\bar{E}_{D_s(1,\frac{1}{2},0)}$ with the
mass of  $D_s(2317)$ we
 find $\bar{E}_{B_s(1,\frac{1}{2},0)}=5634$ MeV. With this energy level
 we can then compute the levels of other P-wave states.
 The result is summarized in 
 Table ~\ref{b-mesons}.
An interesting feature of our estimation is that for $j=1/2$ the $B_s$ mesons have
almost equal or slightly smaller masses than 
their non-strange counterparts.

\begin{table}[ht]
\tbl{Modified energy levels $\bar E$ for P-wave
 charmed mesons.  (Units are in MeV.)}
{\begin{tabular}{@{}ccccccccc@{}}\toprule
               &$D(\frac{1}{2},0)$&$D(\frac{1}{2},1)$&
	        $D(\frac{3}{2},1)$&$D(\frac{3}{2},2)$&
	        $D_s(\frac{1}{2},0)$&$D_s(\frac{1}{2},1)$&
	        $D_s(\frac{3}{2},1)$&$D_s(\frac{3}{2},2)$\\ \hline
$m_{\rm ex.}$&2308&2427&2422&2459&2317&2460&2535&2573 \\
${\bar E} $      &2308&2421&2425&2468&2317&2435&2527&2573 \\ 
$E$            &2377&2490&2417&2460&2487&2605&2535&2581\\
\botrule
\end{tabular}
\label{c-mesons}}
\end{table}
\begin{table}[ht]
\tbl{Modified energy levels $\bar E$ for P-wave
 bottom mesons.  (Units are in MeV.)}
 {\begin{tabular}{@{}ccccccccc@{}}\toprule
 &$B(\frac{1}{2},0)$&$B(\frac{1}{2},1)$&$B(\frac{3}{2},1)$&$B(\frac{3}{2},2)$&
	       $B_s(\frac{1}{2},0)$&$B_s(\frac{1}{2},1)$&$B_s(\frac{3}{2},1)$&
	       $B_s(\frac{3}{2},2)$\\ \hline
$\bar E $      &5637&5673&5709&5723&5634&5672&5798&5813 \\
$E$            &5706&5742&5700&5714&5804&5842&5805&5820\\
\botrule
\end{tabular}
\label{b-mesons}}
\end{table}

\section{Conclusion}
Chiral loop corrections in potential model cannot be ignored since these are
not small, for instance in charmed mesons they are comparable to the $1/M_c$
corrections. However, these were completely ignored in the computations of energy levels
in conventional potential model.
Once the chiral loop corrections are taken into account 
we find important puzzles such as the mass gap and 
spin-orbit inversion can be understood within the potential model.
This suggests that the P-wave states be of more conventional type rather than
exotic states such as four-quark states.

\section*{Acknowledgments}
This work was supported in part by 
Korea Research Foundation 
Grant (KRF-2006-015-C00251)

\end{document}